\documentclass[sigconf]{acmart}
\usepackage{amsmath,nccmath,tabularx} 
\usepackage{algorithm}
\usepackage{algorithmic}
\usepackage{graphicx}
\graphicspath{{images/}}
\usepackage{wrapfig,lipsum}
\usepackage{tabularx}
\usepackage{subcaption}
\usepackage{multirow}
\newcommand\todo[1]{\textcolor{red}{#1}}

\usepackage{enumitem}
\setlist{leftmargin=0mm}
\usepackage[T1]{fontenc}
\usepackage[utf8]{inputenc}
\usepackage[font=small,labelfont=bf]{caption}

\AtBeginDocument{%
  \providecommand\BibTeX{{%
    \normalfont B\kern-0.5em{\scshape i\kern-0.25em b}\kern-0.8em\TeX}}}

\copyrightyear{2022} 
\acmYear{2022} 
\setcopyright{acmlicensed}\acmConference[SIGIR '22]{Proceedings of the 45th International ACM SIGIR Conference on Research and Development in Information Retrieval}{July 11--15, 2022}{Madrid, Spain}
\acmBooktitle{Proceedings of the 45th International ACM SIGIR Conference on Research and Development in Information Retrieval (SIGIR '22), July 11--15, 2022, Madrid, Spain}
\acmPrice{15.00}
\acmDOI{10.1145/3477495.3531951}
\acmISBN{978-1-4503-8732-3/22/07}

\begin{document}
\fancyhead{}
\title{CharacterBERT and Self-Teaching for Improving the Robustness of Dense Retrievers on Queries with Typos}


\author{Shengyao Zhuang}
\affiliation{%
	\institution{The University of Queensland}
	\streetaddress{4072 St Lucia}
	\city{Brisbane}
	\state{QLD}
	\country{Australia}}
\email{s.zhuang@uq.edu.au}

\author{Guido Zuccon}
\affiliation{%
	\institution{The University of Queensland}
	\streetaddress{4072 St Lucia}
	\city{Brisbane}
	\state{QLD}
	\country{Australia}}
\email{g.zuccon@uq.edu.au}


\begin{abstract}
Current dense retrievers are not robust to out-of-domain and outlier queries, i.e. their effectiveness on these queries is much poorer than what one would expect. In this paper, we consider a specific instance of such queries: queries that contain typos. We show that a small character level perturbation in queries (as caused by typos) highly impacts the effectiveness of dense retrievers.
We then demonstrate that the root cause of this resides in the input tokenization strategy employed by BERT. In BERT, tokenization is performed using the BERT's WordPiece tokenizer and we show that a token with a typo will significantly change the token distributions obtained after tokenization. This distribution change translates to changes in the input embeddings passed to the BERT-based query encoder of dense retrievers. We then turn our attention to devising dense retriever methods that are robust to such  queries with typos, while still being as performant as previous methods on queries without typos. For this, we use CharacterBERT as the backbone encoder and an efficient yet effective training method, called Self-Teaching (ST), that distills knowledge from queries without typos into the queries with typos. Experimental results show that CharacterBERT in combination with ST achieves significantly higher effectiveness on queries with typos compared to previous methods. Along with these results and the open-sourced implementation of the methods, we also provide a new passage retrieval dataset consisting of real-world queries with typos and associated relevance assessments on the MS MARCO corpus, thus supporting the research community in the investigation of effective and robust dense retrievers. Code, experimental results and dataset are made available at \url{https://github.com/ielab/CharacterBERT-DR}.

\end{abstract}


\maketitle

\section{Introduction} \label{sec:intro}

Neural ranking models have shown to provide remarkable effective\-ness improvements compared to traditional information retrieval (IR) methods~\cite{lin2021pretrained}. These ranking models rely on heavily pre-trained deep language models, such as BERT~\cite{devlin2019bert} and RoBERTa~\cite{liu2019roberta}. Notably, increasing effort has been devoted to developing effective dense retrievers (DRs)~\cite{karpukhin-etal-2020-dense,xiong2020approximate,gao2021complementing,zhan2020repbert,khattab2020colbert,DBLP:conf/emnlp/GaoC21,gao2021unsupervised,Zhan2021OptimizingDR,ren2021pair,ren2021rocketqav2,qu2021rocketqa}.
 DRs use BERT to encode both queries and documents into low dimensional dense vectors. Relevance between queries and documents is esti\-ma\-ted by similarity matching methods (e.g., cosine similarity) between the dense vectors. Prior work has shown that DRs are much more effective than traditional bag-of-words retrieval methods. Arguably, this is due to a DRs' ability to address the vocabulary mismatch problem: i.e., when queries and the corresponding relevant do\-cu\-ments use different but semantically equivalent terms.


Recent studies however have also highlighted problems with current DRs that relate to their robustness to out-of-distribution queries, that is, queries that have uncommon traits compared to those that form the bulk of data used at training and for which DRs obtain unexpectedly low effectiveness compared to in-distribution queries. For example, \citeauthor{sciavolino2021simple}~\cite{sciavolino2021simple} have shown DRs have poor effectiveness on queries that contain entities, while Arabzadeh et al.~\cite{arabzadeh2021ms} showed poor effectiveness on queries with uncommon terms or complex queries. Another type of queries for which DRs are not robust, and that are the focus of this paper, are queries that contain typos~\cite{zhuang2021dealing,penha2021evaluating,wu2021neural,arabzadeh2021ms}.
\citeauthor{zhuang2021dealing}~\cite{zhuang2021dealing} systematically simulated a wide range of typos in English queries and found that just injecting one character-level typo into MS MARCO queries  dramatically decreases the effectiveness of DRs. Interestingly, they also found that DRs are more sensitive to typos in queries than bag-of-words methods (BM25). They further suggested that queries containing typos give rise to a specific type of vocabulary mismatch and questioned the ability of DRs to deal with this problem.

\begin{figure*}
	\includegraphics[width=\linewidth]{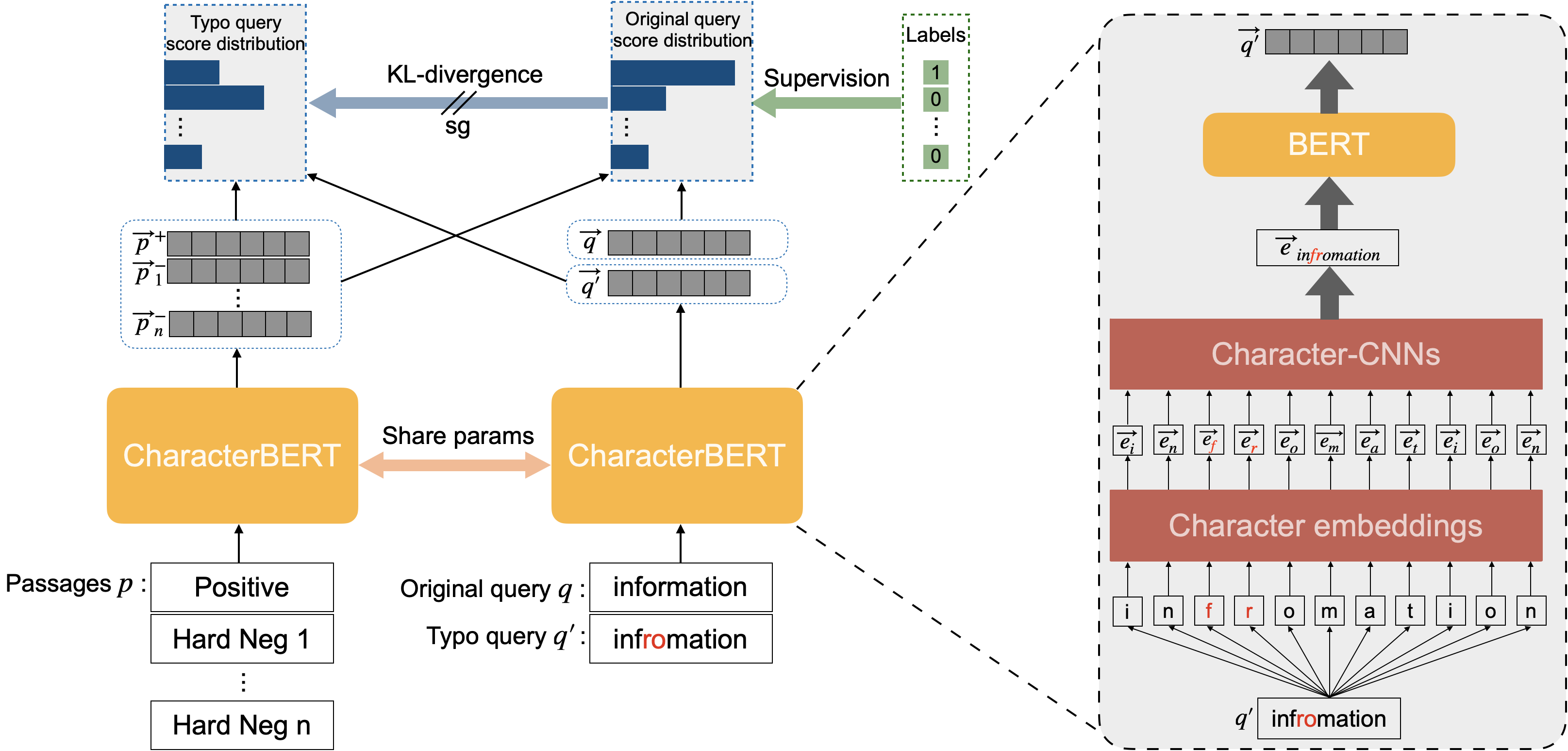}
	\vspace{4pt}
	\caption{Our CharacterBERT + Self-Teaching training approach.}
	\label{fig:1}
\end{figure*}

In this paper, we revisit the DRs' lack of robustness on queries with typos by providing an in-depth analysis of this problem, including unveiling why this occurs, and a concrete solution for addressing this problem in DRs.
Specifically, we demonstrate that DRs are not robust to typos in queries because of the limitations of the BERT's WorldPiece tokenizer~\cite{wu2016google}: a token with a typo will significantly change the token distributions obtained after tokenization -- this dramatically changes the token embeddings used as input to the DRs query encoder, in turn resulting in a very different query embedding than the original one (obtained if typos were not present).
For example, the word ``information'' is mapped to a single token by the BERT tokenizer. Thus, the input passed to the DRs' query encoder for this word is a single embedding.
However, if this word is misspelled into ``inf\todo{ro}mation'' it is then split by the tokenizer into four sub-tokens: [`in', `fr', `oma', `tion']. As a consequence, the input passed to the DR's query encoder for this word is composed of four token embeddings (also with additional position embeddings), which are very different from the single embedding for ``information''. Then, the encoding produced by the DRs' query encoder for the query ``inf\todo{ro}mation'' is understandably different from that for ``information'' -- and empirically this is shown to significantly affect effectiveness.



To overcome this robustness issue when answering queries with typos and based on the above observation, we create a Cha\-rac\-terBERT~\cite{el2020characterbert} DR based query and passage encoder.
Cha\-rac\-terBERT is a variant of BERT in which the WordPiece tokenizer is replaced by a Character-CNN~\cite{peters2018deep,jozefowicz2016exploring} module to construct the embeddings for the tokens' characters. Empirically, we show that simply replacing BERT-based DR encoders with CharacterBERT-based ones can improve the robustness of the DRs on queries with typos (while not hurting the effectiveness on queries without typos). However, this simply strategy still is not enough to bridge the gap in effectiveness between queries with and without typos: effectiveness differences are still considerably large. 
To further improve the effectiveness of DRs on queries with typos, we also propose a novel adversarial training method for DRs called Self-Teaching (ST), illustrated in Figure~\ref{fig:1}. 
Our ST method takes inspiration from methods in Knowledge Distillation (KD)~\cite{hinton2015distilling}. ST distills the model's knowledge constructed when training on queries without typos into the model's knowledge for queries with typos.
The fundamental difference between standard KD training and our proposed ST training is that the student model in our ST training is at the same time also the teacher model. 
This gives our ST training a key advantage: there is no need to have a well trained teacher model in advance or jointly train two different student and teacher models, thus resulting in savings of computational resources.

Our empirical evaluation, that builds upon \citeauthor{zhuang2021dealing}'s synthetic typo query generation ~\cite{zhuang2021dealing}, 
shows that dense retrievers trained with Cha\-rac\-terBERT + ST significantly outperform baseline DRs trained with standard BERT and an augmentation-based training method (pro\-po\-sed in~\cite{zhuang2021dealing}).
In addition to the synthetic typo query evaluation, we also compile a new dataset, DL-typo, with human relevance judgements and characterised by the availability of query pairs (without and with typos),
where queries with typos are not synthetically produced but are real queries mined from a large search engine query log. This new dataset contributes real-world data related to queries with typos that researchers can use to investigate the robustness of dense retrievers. 

\textbf{\textit{Novel Contributions.}} 
We make the following contributions:
\begin{enumerate}
	\item We provide a thorough analysis of the behaviour of BERT-based dense retrievers  on queries with typos. We then identify the reasons behind this behaviour, and in particular the loss in effectiveness.
	
	\item We propose to use CharacterBERT as the backbone of the  dense retrievers' bi-encoder in a bid to improve effectiveness on queries that contain typos. 
	
	
	\item We propose a new Self-Teaching adversarial training regime that distillates the score distribution of queries that do not contain typos to the queries within typos: this is aimed at further improving the effectiveness of dense retrievers on queries with typos.
	
	
	\item We create a new dataset based on MS MARCO passage ranking data that contains 60 query pairs and corresponding relevance assessments. Each pair is constituted by a real query with typos extracted from a large search engine query log and the corresponding query where the typos have been corrected.
	
	\item We thoroughly evaluate the proposed methods on an array of datasets for passage ranking, including datasets where typos are synthetically generated in a controlled manner, and our dataset with real queries with typos. The experimental results suggest that our CharacterBERT-based dense retriever with Self-Teaching  outperforms dense retrievers methods to deal with queries with typos.
	
	

	\item We further compare the effectiveness of the proposed methods for dense retrievers against state-of-the-art spelling correction methods (overlooked in previous work on dense retrievers). These results help contextualise the proposed dense retriever solutions, tease out their advantages and limitations compared to alternative pipelines with spell-checkers, and chart directions for future work.


\end{enumerate}

\vspace{-6pt}
\section{Related Works} \label{sec:related_work}

\textbf{\textit{Effectiveness of Dense Retrievers.}} Pre-trained transformer-based language models (PLMs)~\cite{devlin2019bert,liu2019roberta,yang2019xlnet} from NLP have been adapted to IR tasks such as passage retrieval and have demonstrated much higher effectiveness than traditional bag-of-words methods (e.g., BM25)~\cite{lin2021pretrained}. 
Among the retrieval methods that exploit PLMs, dense retrievers have attracted the most attention. \citet{karpukhin-etal-2020-dense} first demonstrated that simply leveraging hard negatives sampled from BM25 top retrieved passages is sufficient for learning a fairly effective BERT-based DR. 
Follow up work has focused on designing more complex hard negative mining strategies to further improve  effectiveness~\cite{xiong2020approximate,Zhan2021OptimizingDR,qu2021rocketqa}. \citet{xiong2020approximate} proposed to sample hard negatives from an asynchronously updated Approximate Nearest Neighbor (ANN) index, giving rise to the ANCE model. 
Similar to ANCE, \citet{Zhan2021OptimizingDR} suggested that static hard negative sampling is risky and leads to learning a suboptimal DR.  They then proposed to dynamically update the query encoder so that hard negatives are sampled according to the latest query embeddings: this is the ADORE method. 
Training dense retrievers with hard negatives is however computationally expensive. 
To maximise the GPUs computational power, \citet{qu2021rocketqa} proposed to leverages negative samples across different GPUs, thus greatly enlarging the negatives samples per training data while maintaining training efficiency as GPU memory locality is exploited: this is the training pipeline of the RocketQA model. The availability of such a large number of negative samples significantly improves ranking effectiveness. 
However, this training pipeline still requires a large number of GPUs and memory.
As an alternative, \citet{gao2021scaling} introduced a gradient caching technique that allows the RocketQA training pipeline to be run within a single GPU. 
 In this paper, we adopt the hard negative training approach of~\citet{karpukhin-etal-2020-dense} to efficiently train an effective dense retriever, and leave the application of our methods on more complex dense retrievers training pipelines to future work.

Another widely used technique used to improve the effectiveness of dense retrievers is Knowledge Distillation (KD)~\cite{hofstatter2020improving,lin2020distilling,lin-etal-2021-batch,ren2021rocketqav2,izacard2020distilling}, where a DR is obtained by learning from a stronger teacher ranker. 
\citet{lin2020distilling,lin-etal-2021-batch} showed that an effective single vector-based dense retriever can be learnt from the late-interaction ColBERT teacher model. Others have suggested distilling knowledge from expensive but more effective cross-encoder-based rerankers to yield effective bi-encoder-based DRs~\cite{hofstatter2020improving,izacard2020distilling}. On the other hand, \citet{ren2021rocketqav2} showed that a bi-encoder-based student retriever and a cross-encoder-based re-ranker teacher model can be trained jointly. 
Our proposed Self-Teaching training is also inspired by KD but is fundamentally different from KD because it is specifically designed for improving the robustness of dense retrievers to queries with typos and it does not require a specific teacher model, in the sense that the student model itself is also the teacher model.
\textbf{\textit{Robustness of Dense Retrievers.}}
While generally effective, recent research has shown dense retrievers can exhibit poor effectiveness in specific circumstances, thus detracting from their robustness.  
An analysis of the MS MARCO passage ranking leaderboard results has identified that papular DRs perform poorly on queries with uncommon terms and on complex queries (i.e. those that ``require interpretation [...] beyond the immediate meaning of the terms''~\cite{arabzadeh2021ms} in the queries).
Coincidentally, \citet{mackie2021how} introduced a framework for identifying hard queries that challenge DRs. 
\citet{sciavolino2021simple} found that DRs also perform poorly, and worse than traditional bag-of-words methods, on queries that contain entities.
On the other hand, \citet{zhuang2021dealing} investigated the effectiveness of DRs on queries with typos. We build directly upon their findings, and thus we detail their work further.
They proposed a synthetic typo generation framework that given a query without typos it produces a corresponding query that contains a realistic typo generated accordingly to a set of rules derived from common types of misspellings found in search logs for English queries. Using these queries they evaluated the drop in effectiveness caused when dense retrievers have to answer a query with typos, rather than the corresponding original (typo-free) query. 
They found that BERT-based DRs are extremely sensitive to typos: the injection of a single character-level typo into queries dramatically decreases their effectiveness, rendering dense retrievers not robust to typos in queries. We further highlight that research in NLP has also shown that BERT-based methods are not robust to typos~\cite{sun2020adv}, although this research has not considered DRs and the retrieval task.



While previous research has highlighted limitations in terms of robustness, little work has been done towards strengthening dense retrievers and address the underlying problems -- noteworthy exceptions are described next. \citet{chen2021salient} tackled the robustness of DRs on entity queries and proposed a contrastive training mechanism to learn a DR by imitating the behaviour of a sparse retriever behaviour; the results from this are further fused with those from a standard DR to achieve satisfactory effectiveness on both general queries and queries with entities. A drawback of their method is that it requires large quantities of training data and computational resources.
The aforementioned study of \citet{zhuang2021dealing} has investigated another type of queries that challenges the robustness of DRs: queries that contain typos. To address this, they devised an augmentation-based adversarial training method for dense retrievers (called typo-aware training) which, with a set probability, transforms training queries that do not contain typos into queries with typos.
Their results show that DRs trained with their augmentation technique are more robust to queries with typos; furthermore, these same models do not show losses in effectiveness when used on queries that do not contain typos. 
That work however had two limitations: (i) it did not provide insights on why DRs are not effective on queries with typos, and (ii) it only evaluated methods on realistic but synthetically generated typos in queries. 
Our work builds upon this and it contributes the following new knowledge and resources: (1) an in-depth analysis of the behaviour of BERT-based DRs on queries with typos, (2) insights into why DRs are sensitive to queries with typos, (3) a new training method to further improve the robustness of DRs on typo queries, (4) a new dataset with human relevance judgements and real queries with typos to evaluate the robustness of DRs.

\begin{table}[t]\small
	\begin{tabular}{|l|c|c|c|c|c|c|c|c|}
		\hline
		Diff  & 1   & 2    & 3    & 4   & 5  & 6 & 7 & Total \\ \hline
		Count & 803.9 & 2,926.6 & 2,431.6 & 712.4 & 91.5 & 8.2 & 0.8 & 6,975  \\ \hline
	\end{tabular}
	\caption{Average distribution of \textit{tokenization difference} for 10 replicas of the MS MARCO dev dataset.}
	\label{table:1}
\end{table}

\section{Impact of typos on tokenization and query embeddings}\label{sec:3}

In this section we investigate the impact of a typo in a query on dense retrievers, in terms of both representation and effectiveness: this provides the motivations for investigating methods to make dense retrievers robust to typos and for the proposed use of CharacterBERT and ST, along with a clear understanding of why dense retrievers are not effective when queries have typos.

For analysing this problem we rely on the synthetic typo generation process proposed by~\citet{zhuang2021dealing} where, for all queries in MS MARCO dev, typos are randomly inserted in a query token using 5 typo generators: RandInsert, RandDelete, RandSub, SwapNeighbor, and SwapAdjacent\footnote{See Section 2.1 of~\citet{zhuang2021dealing} for details about these typo generators}.
Note the method only inserts one typo per query, i.e. only one word in the query contains a typo, and only words with at least 3 characters and not contained in a standard stopword list are considered as candidates for typo-insertion\footnote{Because of these constrains, 5 queries out of 6,980 have been discarded.}.
This procedure then generates a dataset based on the MS MARCO dev query set which contains 6,975 pairs of queries: each pair is composed of the original MS MARCO query and the corresponding query with a typo.
We repeat this typo generation process 10 times, so that results are less likely to be influenced by the specific tokens chosen for typo-insertion and the type of typos. Results presented next are averaged across the 10 replicas.


We first investigate the consequence a typo in the query has on the behaviour of the BERT WordPiece tokenization. 
Table~\ref{table:1} reports how many tokens are different between each pair of original query and query with typo.  So, for instance, on average across the 10 replicas of this synthetic dataset, for 2,926.6 queries the BERT WordPiece tokenizer produces tokenizations of the queries with typos that differ from the original queries tokenizations for 3 tokens: we call this a \textit{tokenization difference} of 3. In other words, the tokenization for the query with typo has 3 tokens that are different compared to the tokens obtained from the tokenization of the original query. In practice, this often also implies that the tokenizations of queries with typos have more tokens than the corresponding tokenization of the original queries without typos, but this is not always the case. For example if the original query contains ``apple'' and the corresponding query with typo has changed this word into ``apply'', then the BERT WordPiece tokenizer will output one token for each of these words (i.e. their tokenizations have the same length), but the obtained tokens are different and thus the two queries have a tokenization difference of 1.

From the results in Table~\ref{table:1} we note that the majority of queries with typos result in tokenization differences of 2 or 3  tokens. In some rare cases, a typo in a query can even result in a tokenization difference of up to 6 or 7 tokens -- importantly, however, no query pair leads to a tokenization difference of zero. Given that the BERT WordPiece tokenizer produces tokenizations for the original MS MARCO dev queries that contain on average 7 tokens, it is self-evident that even a small tokenization difference corresponds to sensibly different queries being then encoded by the DRs: a difference of 2-3 tokens corresponds to 29\%-/43\% of the tokens being different. Thus, it is not surprising that in practice BERT-based DRs that rely on the BERT WordPiece tokenizer are extremely sensitive to typos in  queries, as these typos can largely change the input tokens passed to the DRs query encoders.

Next we investigate the differences in the representations produced by dense retrievers for original queries and queries with typos, and how these differences in representations lead to differences (and specifically drops) in effectiveness. To measure differences in representations, we measure the cosine similarity between the DR encoding of the original query and the DR encoding of the corresponding query with typos: the smaller the cosine, the higher the difference. We call this \textit{encoding similarity}. To measure differences in effectiveness, we measure the \textit{MRR drop rate} ($\Delta_{MRR}$), i.e., the average RR score decrease rate between original queries ($q_i$) and queries with typos ($q'_i$): $\Delta_{MRR} = \frac{\sum_{\forall <q_i,q'_i>}  RR(q_i) - RR(q'_i)}{\sum_{\forall <q_i,q'_i>} RR}$.
For these experiments, we use as representative method our baseline DR model (labelled \texttt{StandardBERT-DR}) trained on the MS MARCO training data with BM25-based hard negative sampling (see Section~\ref{sec:baselines} for details on this baseline). Other DRs give rise to similar observations and are omitted here for clarity.

Results are reported in Figure~\ref{fig:droprate-to-cosine} (solid lines), where differences in representations and differences in effectiveness are analysed across values of tokenization difference (ignore the dashed lines -- these are analysed in Section~\ref{sec:results}). First, we observe that values of $\Delta_{MRR}$ are sensibly large, i.e., queries with typos are less effective than the corresponding queries without typos. This is in line with previous work~\cite{zhuang2021dealing,penha2021evaluating,wu2021neural} and occurs regardless of the value of tokenization difference. However, as tokenization difference increases, so does $\Delta_{MRR}$: queries with typos for which their tokenized version largely differs from that of the corresponding original query lead to larger losses in effectiveness. We also note that as $\Delta_{MRR}$ and tokenization differences grow hand-in-hand, encoding similarity instead behaves exactly in the opposite way: it decreases as tokenization differences increases. In other words: the less the injected typo affects the BERT WordPiece tokenizer, the less tokenization difference is, and thus the more similar the encodings produced for the original and the query with typo are, resulting in smaller differences (losses) in effectiveness between original and typo query -- and vice-versa when the injected typos have a high effect on the BERT WordPiece tokenizer. These provide us with the following intuition and the motivations for the methods we proposed in Section~\ref{sec:approch}: a typo-robust dense retriever needs to produce query encodings that are invariant to the presence of typos in queries.

\section{Typo-robust dense retrieval with CharacterBERT and Self-teaching} \label{sec:approch}

Section~\ref{sec:3} outlined the intuition that typo-robust dense retrievers require tokenization and query encoding mechanisms that are not sensitive to typos: in particular, the query encodings should be \textit{invariant} to the presence or absence of typos in the query\footnote{i.e., the encoding of the query with typos should be the same, or have minimal differences compared to the encoding of the corresponding query without typo.}.
Next, we build upon this intuition and propose to use CharacterBERT, which uses a typo insensitive tokenization approach, as encoder in DRs; in addition we propose an efficient Self-Teaching training approach for DRs that specifically aims to render the query encoder  invariant to the presence of typos in the query.


\subsection{Typo-Robust DRs with CharacterBERT}

The CharacterBERT model uses a Character-CNN module to construct the word embeddings that are then passed as input to BERT~\cite{el2020characterbert}, thus dropping the need for the WordPiece tokenizer used in standard BERT models.
The architecture outlined in Figure~\ref{fig:1} shows that any word passed as input to CharacterBERT is first split into its characters.
Each character then is used to retrieve the corresponding character embedding from a character embedding matrix; the character embedding is then passed as input to a stack of CNN layers.
 The outputs of the CNN layers are then aggregated and projected into a single vector representation for each token. These new token representations can serve as context-independent word embeddings which can therefore be combined with position and segment embeddings before being fed into BERT.


The key difference between BERT and CharacterBERT is that, instead of using a set of pre-trained token embeddings obtained from the WordPiece tokenizer, CharacterBERT produces a single embedding for \textit{any} input word without relying on any pre-defined vocabulary (as WordPiece instead does).
 This is particularly desired for our task since with CharacterBERT a query word with a typo will then be represented by a single embedding. This is in contrast to the standard BERT models that rely on the WordPiece tokenizer, where a query word with a typo will often be represented by multiple token embeddings. Thus, a dense retriever encoder implemented with CharacterBERT may be more robust to typos in queries. This expectation is derived also from the fact that CharacterBERT has been found to be more robust than BERT to out-of-vocabulary words across a variety of NLP tasks, in particular in specialized domains~\cite{el2020characterbert}. However, no previous work has investigated how CharacterBERT could help tackle the problem of robustness of dense retrievers when typos are in queries (and no previous work has used CharacterBERT to build a neural ranker). 


We propose to adapt CharacterBERT to obtain a typo-robust dense retriever. Specifically, given a traditional bi-encoder DR architecture, we modify it so as to use the [CLS] token embedding output from CharacterBERT to encode both queries and passages  into a single vector each, as opposed to relying on BERT for this:


\begin{equation}
CharacterBERT(q) = \vec{q},  \mbox{\hspace{10pt}} CharacterBERT(p) = \vec{p}, 
\end{equation}
\noindent where $\vec{q}$ and $\vec{p}$ are the encoded versions of query $q$ and passage $p$. Note that the reliance on the bi-encoder architecture means passages can be encoded offline at indexing and at query time only the query requires encoding, thus retaining the efficiency advantage that characterises DRs. The ranking results are then constructed by Approximate Nearest Neighbor (ANN) search, where the similarity between query $q$ and passage $p$ is defined using the dot product: 


\begin{equation}
s(q,p) = CharacterBERT(q)^T \cdot CharacterBERT(p) = \vec{q}^T\cdot\vec{p}
\end{equation}

The proposed method aims to address the impact of typos contained in queries. We do this by encoding queries and documents using CharacterBERT. An alternative approach would have been to only encode queries  using CharacterBERT, as queries are the items that are likely to contain typos, and maintain the regular BERT encoder for encoding passages. This renders the architecture at all effects an `untied' bi-encoder, i.e. a bi-encoder where the query and the document encoder do not share parameters -- and in our case are at all effects different encoding methods. 
However, in our initial experiments, we found that this `untied' architecture performs significantly worse than just using a single CharacterBERT to encode both queries and passages. Our hypothesis for this behaviour is that CharacterBERT may tend to generate  query encodings that are very different to those of the regular BERT, making it harder to train both encoders to generate similar vectors for relevant query-passage pairs.


\subsection{Knowledge Distillation with Self-Teaching}
Knowledge distillation (KD) has become a common strategy for training effective dense retrievers: when using KD, a student DR model can learn from a stronger multi-vector DR teacher (like ColBERT)~\cite{lin2020distilling} or from a cross-encoder teacher~\cite{hofstatter2020improving,lin-etal-2021-batch,ren2021rocketqav2}. 
However, these KD training methods often require large computational resources because either a well trained teacher model with fixed parameters is needed, or jointly training of both student and teacher models is required.
On the contrary, we design a novel KD training approach to learn typo-robust DRs that goes beyond such a computational requirement limitation: in our method, the model itself is both the student and the teacher and thus only one model is optimized during training. We call this approach \textit{Self-Teaching (ST)}.


The design of our Self-Teaching method is based on the following intuition. A key observation emerges from the findings in Figure~\ref{fig:droprate-to-cosine} and previous works~\cite{zhuang2021dealing,penha2021evaluating,wu2021neural}: Dense retrievers have much higher effectiveness on a query without typos than on its corresponding version with typos.
This implies that dense retrievers can effectively ``understand'' and model (encode) the query, thus performing a better job in terms of relevance estimation, if the query itself does not contain typos. Hence, to improve the effectiveness of dense retrievers on queries with typos we can inject the dense retriever knowledge learnt from the corresponding query without typos.



Specifically, the proposed ST training applies one of many transformations to each query $q$ in the training query set $\mathcal{Q}$. Such transformations mimic the common typos that users make in queries: for each query $q$ a transformation is randomly picked\footnote{In our experiments, each transformation has a uniform probability of being picked; a different distribution may be chosen to mimic popularity of types of typos, e.g., as observed from query logs for the specific search service.} from the set of available transformations and applied to a random word (with constrains on the selection of such word) in the query, to generate a query $q'$ with a typo. In terms of transformations, in our experiments we use the typo generators described in Section~\ref{sec:3} and also used in previous work~\cite{zhuang2021dealing}. We note realistic typo generations are language dependent; those we consider are realistic for English. The pair $<q,q'>$ shares the same list of candidate passages $\mathcal{P}_q = \{p_{q}^i\}_{1\leq i	\leq m}$ that are associated with the original query q. Then, the similarity score between $q$ and any passage $p_{q}^i$ can be computed as $s_q(q,p_{q}^i)$; similarly $s_{q'}(q',p_{q}^i)$ can be computed for  $q'$. These scores are then normalised via Softmax:


\begin{equation}
 \tilde{s}_q(q,p) = \frac{e^{s_q(q,p)}}{\sum_{p_{q}^i\in{\mathcal{P}_q}} e^{s_q(q,p_{q}^i)}}, \; \tilde{s}_{q'}(q',p) = \frac{e^{s_{q'}(q',p)}}{\sum_{p_{q}^i\in{\mathcal{P}_q}} e^{s_{q'}(q',p_{q}^i)}}
\end{equation}

The goal of our ST training is to minimise the difference between the score distribution obtained from the query with the typo ($q'$) and the score distribution obtained from the corresponding query without typo ($q$, the original query). For this, we minimize the KL-divergence loss:

\begin{equation}
\mathcal{L}_{KL} (\tilde{s}_{q'}, \tilde{s}_q) =  \tilde{s}_{q'}(q',p) \cdot \log \frac{\tilde{s}_{q'}(q',p)}{ \tilde{s}_q(q,p)}
\end{equation}

This KL-divergence loss itself does not provide any relevance signal: it only reduces the score difference obtained by the original query and the query with the typo. Thus, we also provide ground-truth relevance labels to supervise the DR learning process so that it can learn the relevance relationships between queries and passages. 
Specifically, the candidate passages set $\mathcal{P}_q$ of each query for the MS MARCO dataset\footnote{Which is used in this paper to train the dense retrievers.} contains on average one positive passage $p^+$ and a set of negative passages $\{p_{1}^-, p_{2}^-, ..., p_{m-1}^-\}$; these labelled passages are used to train with a supervised contrastive cross-entropy loss:


\begin{equation}
\mathcal{L}_{CE} ({s}_{q})= -\log \frac{e^{s_{q}(q,p^+)}}{e^{s_{q}(q,p^+)} + \sum_{p^-}e^{s_{q}(q,p^-)}}
\end{equation}

Finally, we combine the KL loss and supervised contrastive loss to form our ST loss $\mathcal{L}_{ST} = \mathcal{L}_{CE} ({s}_{q}) + \mathcal{L}_{KL} (\tilde{s}_{q'}, sg(\tilde{s}_q))$, where the $sg(.)$ function is used to stop the gradients from the score distribution computed for the original query (without typos) in the KL loss. We find that this gives rise to a more stable training and yields slightly higher overall effectiveness. 

\section{Experimental settings} \label{sec:experimental}

\subsection{Baselines and Implementations} \label{sec:baselines}
The goal of the methods proposed in this paper, namely Cha\-ract\-er\-BERT-based encoders and Self-Teaching training, is to produce dense retrievers that are as effective as standard dense retrievers on queries without typos, but are far more effective than these on queries that contain typos. To evaluate whether this goal is achieved, we then compare the proposed methods against popular and effective DRs that do not explicitly tackle the problem of queries with typos: ANCE~\cite{xiong2020approximate} and TCT-ColBERTv2~\cite{lin-etal-2021-batch}. The ANCE method is characterised by hard negatives sampled in an asynchronously iterative way as the DR is trained, and this is combined with a corresponding update of the ANN index. The TCT-ColBERTv2 method is characterised by the learning from a strong cross-encoder teacher model.
Both methods are expensive to train and require high computational resources. Thus, we directly use the corresponding implementations,  model checkpoints and pre-built ANN indexes from Pyserini~\cite{Lin2021pyserini}; Pyserini is also used to produce baseline BM25 results on all datasets, for completeness.

We highlight that our proposed CharacterBERT and ST methods can be applied on top of any DR, and thus also in combination with the selected benchmark methods ANCE and TCT-ColBERTv2: this would make these benchmark models robust to queries with typos. However, in our experiments we do not do this because of the aforementioned high training costs associated to these DRs  -- we leave these comparisons to future work. 

Finally, we note that ANCE and TCT-ColBERTv2 have a number of carefully designed training ``tricks'', specifically aimed at gaining maximum effectiveness on standard (without typo) queries. 
Because of this, we also implement a vanilla standard BERT dense retriever (\texttt{StandardBERT-DR}), which to large extents resembles the DPR approach~\cite{karpukhin-etal-2020-dense} (but performs better overall). \texttt{StandardBERT-DR} is then directly comparable to our CharacterBERT dense retriever (\texttt{CharacterBERT-DR}), which is trained in a similar manner, aside from the two architectures differing from their use of BERT and the WordPiece tokenizer for \texttt{StandardBERT-DR} vs. the only CharacterBERT model for the \texttt{CharacterBERT-DR}. For both models, we follow the hard negative training of DPR and use Tevatron DR training toolkit~\cite{Gao2022TevatronAE} to train from scratch and use the Asyncval toolkit~\cite{zhuang2022asyncval} to validate model checkpoints during training.
For each query in the MS MARCO training set, we randomly sample 7 hard negative passages from the top 200 passages retrieved by BM25 and one positive passage from the qrels.
We set the batch size to 16 and apply in-batch negatives sampling to each training sample in the batch, resulting in $7 + 8 * 15 = 127$ negatives per training sample. We train with the AdamW optimizer and a 5e-6 learning rate, linear learning
rate schedule for 150,000 updates. 
The training of a model can be finished in about a day on a single Tesla V100 32G GPU.
In terms of pre-trained language models, we use the BERT-base-uncased checkpoint provided by the Huggingface transformers library~\cite{wolf-etal-2020-transformers} for \texttt{StandardBERT-DR}, and the CharacterBERT checkpoint  pre-trained on general domain data obtained from the original CharacterBERT repository\footnote{https://github.com/helboukkouri/character-bert} for \texttt{CharacterBERT-DR}.

In addition, we also compare our methods against the only other method that has been proposed in the literature to deal with queries with typos in the context of dense retrievers, namely the typos-aware training approach~\cite{zhuang2021dealing}, which we label with \texttt{+ Aug} to indicate that it represents a data augmentation method. This method uses realistic typo generators to augment training queries and train standard BERT-based DRs to be typo-robust. Thus, in the experiments, we train \texttt{StandardBERT-DR} and \texttt{CharacterBERT-DR} under the following conditions:
\begin{itemize}
	\item \texttt{StandardBERT-DR+Aug}: the original method by ~\citet{zhuang2021dealing};
	\item \texttt{CharacterBERT-DR+Aug}: this relies on our proposed CharacterBERT based dense retriever, and uses the augmentation-based typo-aware training approach by ~\citet{zhuang2021dealing};
	\item \texttt{StandardBERT-DR+ST}: the BERT-based dense retriever is trained using our proposed Self-Teaching method: thus is it similar to the \texttt{StandardBERT-DR+Aug} where the augmentation-based typo-aware training approach is replaced by Self-Teaching
	\item \texttt{CharacterBERT-DR+ST}: this relies on our proposed CharacterBERT based dense retriever and our Self-Teaching method.
\end{itemize}

\subsection{Datasets and Evaluation}
\textit{\textbf{Datasets.}} We employ four datasets to evaluate the proposed methods. 
These are the dev query set of MS MARCO v1 passage ranking dataset~\cite{nguyen2016ms}, the TREC Deep Learning Track Passage Retrieval Task 2019~\cite{craswell2019overview} (DL 2019) and 2020~\cite{craswell2020overview} (DL 2020), and a dataset we compile for this paper (DL-typo). All datasets use the 8.8 million passages released with  MS MARCO, and differ in terms of the queries used. 

The dev query set of MS MARCO contains 6,980 queries, each with on average one relevant passage per query. DL2019 and DL2020 instead contain 43 and 54 judged queries (from the MS MARCO dev set) but with more complete relevance assessments: on average 215.3 and 210.9 judged passages respectively (they may be relevant or not). For DL2019 and 2020, relevance judgements are graded and range from 0 (not relevant) to 3 (highly relevant). Relevance label 1 indicates passages on-topic but not relevant and hence we conflate these passages to label 0 when computing binary metrics (e.g., MAP, MRR), as per standard practice with these datasets. 
For these three datasets, we apply the synthetic typo query generation process~\cite{zhuang2021dealing} outlined in Section~\ref{sec:3}: for each query in a dataset, we apply one of a set of transformations that introduce a typo in the query, and to this query with typo we assign the same relevance assessments of the corresponding query without typo. We repeat the typo generation process 10 times for each dataset (i.e. each original query gives rise to 10 variations with typos) to evaluate average effectiveness of models on typo queries, across different types of typos and different query words being affected by typos.

\begin{table}[t]\small
	\resizebox{\columnwidth}{!}{
		\begin{tabular}{|c|ccccc|}
			\hline
			\multirow{2}{*}{\#Queries} & \multicolumn{5}{c|}{Average number of assessed passages per query, wr.t. relevance label}                                                                            \\ \cline{2-6} 
			& \multicolumn{1}{c|}{0 (Irrelevant)} & \multicolumn{1}{c|}{1 (Related)} & \multicolumn{1}{c|}{2 (Relevant)} & \multicolumn{1}{c|}{3 (Perfect)} & Total \\ \cline{1-1} 
			60                       & \multicolumn{1}{c|}{37.75}          & \multicolumn{1}{c|}{11.50}       & \multicolumn{1}{c|}{9.67}         & \multicolumn{1}{c|}{4.60}        & 63.52 \\ \hline
		\end{tabular}
	}
	\caption{Statistics of our DL-typo dataset.}
	\label{table:2}\vspace{-15pt}
\end{table}

We also compile a purposely built dataset, called \textit{DL-typo}, to analyse effectiveness on real queries with typos. 
For this,  we sample pairs of queries from a public large-scale query spelling correction corpus released by~\citet{hagen2017large}: each pair is composed of a query with typo and its corrected version (by human annotators).  
These queries with typos come from the anonymized AOL query log~\cite{pass2006picture}. 
From this corpus we evenly sample 60 queries with typos (and the corresponding corrected queries) across four common typo types, i.e., 15 RandInsert, 15 RandDelete, 15 SwapNeighbor and 15 RandSub (including 3 SwapAdjacent). 

We then contribute relevance assessments against passages in MS MARCO. To decide which passages to assess for relevance, we issue the 60 queries with typos and their corresponding corrected queries (hence 120 queries in total) to 7 retrieval models\footnote{BM25,  \texttt{StandardBERT-DR},  \texttt{StandardBERT-DR+Aug},  \texttt{StandardBERT-DR+ST},  \texttt{CharacterBERT-DR}, \texttt{CharacterBERT-DR+Aug}, \texttt{CharacterBERT-DR+ST}.} and retrieve passages from MS MARCO. We then pool the top 10 retrieved passages for each run and exhaustively  judge the relevance of the passages in the pool according to the TREC DL juSTs a d guidelines\footnote{https://trec.nist.gov/data/deep2020.html} using Relevation~\cite{koopman2014relevation}. Table~\ref{table:2} reports the key statistics of our DL-typo dataset.


The training queries of the MS MARCO dataset are used for training all the models in this paper: there are about 0.5 million queries in such training dataset, and these were logged by the MS Bing search engine; the training dataset has on average one relevant passage per query (i.e., one positive label per query).

\textit{\textbf{Evaluation Metrics.}} Following common practice, to evaluate the ranking effectiveness of the dense retrievers we use the metrics originally used by the creators of each dataset. For the MS MARCO dev queries dataset, these are Mean Reciprocal Rank at top 10 (MRR@10) and Recall at 1,000 (R@1000). For TREC DL 2019 and 2020, these are Normalized Discounted Cumulative Gain at 10 (nDCG@10), Mean Reciprocal Rank at 1,000 (MRR) and Mean Average Precision (MAP). For these datasets, queries with typos are generated 10 times for each original query without typo: thus for the evaluation on queries with typos we report the metrics averaged for each repeated experiment.
For our DL-typo dataset, since we follow the TREC DL standard assessment practice, we also use the same evaluation metrics used in TREC DL.  Several repetitions of queries with typos are not performed for our dataset because these queries are not synthetically generated. 
Statistical significant differences between methods' results are detected using a two-tailed paired t-test with Bonferroni correction.

\begin{table*}\small
	\resizebox{\textwidth}{!}{
		\begin{tabular}{|c|l|ll|lll|lll|}
			\hline
			\multirow{2}{*}{Queries}                                                    & \multicolumn{1}{c|}{\multirow{2}{*}{Methods}} & \multicolumn{2}{c|}{MS MARCO}                                 & \multicolumn{3}{c|}{TREC DL 2019}                                               & \multicolumn{3}{c|}{TREC DL 2020}                                               \\ \cline{3-10} 
			& \multicolumn{1}{c|}{}                         & \multicolumn{1}{l}{MRR@10} & \multicolumn{1}{l|}{R@1000} & \multicolumn{1}{l}{nDCG@10} & \multicolumn{1}{l}{MRR} & \multicolumn{1}{l|}{MAP} & \multicolumn{1}{l}{nDCG@10} & \multicolumn{1}{l}{MRR} & \multicolumn{1}{l|}{MAP} \\ \hline
			\multirow{9}{*}{\begin{tabular}[c]{@{}c@{}}Without\\ Typos\end{tabular}} 
			& a) \texttt{BM25}                                          & .187                       & .857                             & .497                        & .685                   & .290                     & .487                        & .659                   & .287                     \\
			& b) \texttt{ANCE}~\cite{xiong2020approximate}                                          & .330$^{a}$                    & .959$^{aefghi}$                             & .645$^{a}$                           & .837$^{d}$                   & .371                     & .641$^{ae}$                        & .790$^{a}$                   & .403$^{a}$                     \\
			& c) \texttt{TCT-ColBERTv2}~\cite{lin-etal-2021-batch}                                         & \textbf{.358$^{abdefghi}$}                       & \textbf{.969$^{abdefghi}$}                             & \textbf{.720$^{abdefghi}$}                        & \textbf{.887$^{ade}$}                   & \textbf{.447$^{abdefghi}$}                     & \textbf{.689$^{abdefghi}$}                        & \textbf{.839$^{ae}$}                   & \textbf{.475$^{abdefghi}$}                     \\ 
			& d) \texttt{StandardBERT-DR}                                & .325$^{a}$                       & .953$^{a}$                             & .608$^{a}$                        & .719                   & .353                     & .633$^{ae}$                        & .798$^{a}$                   & .407$^{ag}$                     \\
			& e) \texttt{CharacterBERT-DR}                        & .327$^{a}$                       & .950$^{a}$                             & .609$^{a}$                        & .772                   & .340                     & .586$^{a}$                        & .744                   & .379$^{a}$                     \\ \cline{2-10} 
			& f) \texttt{StandardBERT-DR+Aug}~\cite{zhuang2021dealing}                                & .325$^{a}$                       & .951$^{a}$                             & .620$^{a}$                        & .803                   & .347                     & .629$^{ae}$                        & .794$^{a}$                   & .398$^{a}$                     \\
			& g) \texttt{StandardBERT-DR+ST}                                     & \textbf{.331$^{a}$}                       & .949$^{a}$                             & .616$^{a}$                        & $.826^d$                   & .345                     & .631$^{ae}$                        & .781                   & .393$^{a}$                     \\
			& h) \texttt{CharacterBERT-DR+Aug}                       & \textbf{.331$^{a}$}                       & .949$^{a}$                             & .634$^{a}$                        & $.841^d$                   & .343                     & .612$^{a}$                        & \textbf{.843$^{a}$ }                   & .398 $^{a}$                    \\
			& i) \texttt{CharacterBERT-DR+ST}                            & .325$^{a}$                       & .950$^{a}$                             & \textbf{.643$^{a}$}                        & \textbf{.845$^{de}$}                  & .340                     & .606$^{a}$                        & .827$^{ae}$                    & .390$^{a}$                     \\ \hline \bottomrule
			\multirow{9}{*}{\begin{tabular}[c]{@{}c@{}}With\\ Typos\end{tabular}}     
			& j) \texttt{BM25}                                          & .095                       & .611                             & .256                        & .340                   & .147                     & .291                        & .410                   & .168                     \\
			& k) \texttt{ANCE}~\cite{xiong2020approximate}                                            & \textbf{.200$^{jmn}$}                       & .803$^{jmn}$                             & .448$^{j}$                        & \textbf{.643$^{j}$}                   & .247                     & .461$^{jmn}$                        & .603                   & .278$^{ad}$                     \\
			& l) \texttt{TCT-ColBERTv2}~\cite{lin-etal-2021-batch}                                           & .199$^{jmn}$                       & \textbf{.806$^{jmn}$}                             & \textbf{.449$^{jm}$}                        & .597$^{j}$                   & \textbf{.256$^{jm}$}                     & \textbf{.471$^{ade}$}                        & \textbf{.615}                   & \textbf{.302$^{ad}$}                     \\
			& m) \texttt{StandardBERT-DR}                               & .136$^{j}$                       & .688$^{j}$                             & .298                        & .427                   & .167                     & .323                        & .459                   & .194                     \\ 
			& n) \texttt{CharacterBERT-DR}                        & .159$^{jm}$                        & .724$^{jm}$                              & .355                        & .506                   & .189                     & .326                        & .484                   & .203                     \\ \cline{2-10} 
			& o) \texttt{StandardBERT-DR+Aug} ~\cite{zhuang2021dealing}                                & .215$^{jmn}$                      & .841$^{jklmn}$                            & .434$^{jm}$                       & .582                   & .235$^m$                    & .466$^{jmn}$                       & .630                   & .280$^{ad}$                    \\
			& p) \texttt{StandardBERT-DR+ST}                                     & .228$^{jklmn}$                       & .856$^{jklmn}$                              & .443$^{jm}$                        & .615$^{jm}$                   & .241$^m$                     & .481$^{jmn}$                        & .629                   & .284$^{ad}$                     \\
			& q) \texttt{CharacterBERT-DR+Aug}                       & .251$^{jklmnop}$                       & .877$^{jklmno}$                             & .498$^{jmn}$                        & .688$^{jm}$                   & .264$^{mn}$                     & .486$^{jmn}$                        & .683$^{jmn}$                   & .300$^{ade}$                     \\
			& r) \texttt{CharacterBERT-DR+ST}                            & \textbf{.263$^{jklmnop}$ }                     & \textbf{.894$^{jklmnopq}$}                           & \textbf{.519$^{jmp}$ }                       & \textbf{.706$^{jm}$ }                  & \textbf{.268$^{jmn}$ }                    & \textbf{.514$^{jmn}$}                       & \textbf{.722$^{jmn}$}                  & \textbf{.314$^{jmn}$}                    \\ \hline
		\end{tabular}
	}
	\caption{Results obtained on queries without typos and queries for which typos are obtained synthetically. We repeat the typo generation procedure 10 times, averages are computed first per seed query, and then across queries (distributions for statistical significance computation are formed from the first average). Methods statistically significantly better ($p<0.05$) than others are indicated by superscripts.}
	\label{table:3}
\end{table*}

\begin{figure}
	\includegraphics[width=1\linewidth]{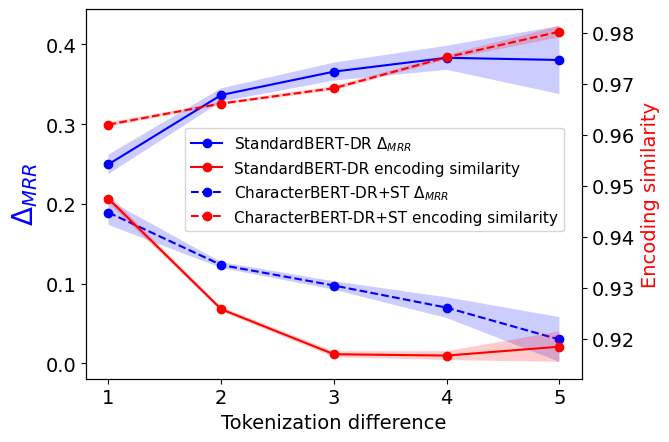}
	\caption{Analysis of MRR drop rate ($\Delta_{MRR}$) and encoding similarity for different values of tokenization difference. Shaded area is standard deviation across queries in the 10 typos variations. Tokenization difference of 6 and 7 are discarded from this analysis as these bins contain too few samples (and thus large variance), although they still fit the trends observed for smaller tokenization difference values. }
	\label{fig:droprate-to-cosine}
\end{figure}

\section{Results} \label{sec:results}
In this section, we first report the results of our proposed methods against baselines on the datasets where synthetic typo generation was used. 
We then present the results obtained on our DL-typo dataset, which reflects the model performance on real user queries with typos.  Finally, we compare our approach with an alternative search engine architecture that involves the use of spell-checkers in the query pre-processing  steps to identify and correct typos: this pipeline is what currently many production search engines rely on for dealing with typos in queries.


\subsection{Results on Queries with Synthetic Typos}

The main results on MS MARCO, TREC DL 2019 and 2020 for both queries with and without typos are reported in Table~\ref{table:3}.

First we consider methods that do not deal with typos in queries (runs a-d in Table~\ref{table:3})  and their effectiveness on queries without typos. We observe that all dense retrievers outperform the bag-of-words baseline (\texttt{BM25}) , as expected. \texttt{ANCE} and \texttt{TCT-ColBERTv2} are more effective than our vanilla \texttt{StandardBERT-DR}:  These results are expected as \texttt{StandardBERT-DR} uses simple hard negative sampling practice while the other two DRs use much more sophisticated, and computationally expensive, training strategies.  Even so, \texttt{StandardBERT-DR} achieves similar MRR@10 and R@1000 as \texttt{ANCE} on MS MARCO dev queries (no statistically significant difference), and higher MAP and MRR on TREC DL 2020 (improvements are however not significant). 

 We now compare \texttt{StandardBERT-DR} and \texttt{CharacterBERT-DR}: these two models are trained with exactly the same settings and negative sampling, and the only difference is the encoder used to encode queries and passages\footnote{While, the results from \texttt{ANCE} and \texttt{TCT-ColBERTv2} are not directly comparable because of the more sophisticated training regimes used by these methods, and are only provided for contextualisation purposes} -- we argue that the encoder of \texttt{CharacterBERT-DR} is more suitable to deal with typos in queries. The two dense retrievers have very similar effectiveness across different metrics and datasets, with the only statistically significant difference obtained on nDCG@10 for TREC DL 2020. These results suggest that, on queries without typos, using CharacterBERT as the core of the dense retrievers' encoders tends to have similar effectiveness as the use of BERT. It is expected, although not empirically proven here, that the adaptation of CharacterBERT to the same training regimes of \texttt{ANCE} and \texttt{TCT-ColBERTv2} would lead to effectiveness similar to these more advance methods: We leave confirming this hypothesis to future work.

We then turn our attention to the results obtained on queries without typos when training with methods that attempt to make DRs more robust on queries with typos -- namely runs f-i, which are based on \texttt{StandardBERT-DR} or \texttt{CharacterBERT-DR} and the two considered training methods \texttt{Aug}~\cite{zhuang2021dealing} and \texttt{ST} (proposed here). We find that these training strategies, although directed at improving effectiveness on queries with typos, often lead to higher effectiveness on queries without typos, regardless on the encoder type used (however, no difference -- be it a gain or a loss -- is statistically significant).
For example, both \texttt{CharacterBERT-DR+Aug} and \texttt{CharacterBERT-DR+ST} achieve higher nDCG@10 and MRR than \texttt{CharacterBERT-DR} on TREC DL 2019 and 2020. The same trend is observed on TREC DL 2019 when analysing the results for \texttt{CharacterBERT-DR}.


Next, we consider queries with typos. The methods that are not explicitly designed to tackle queries with typos, i.e. runs j-m in Table~\ref{table:3}, return results that are dramatically lower than their counterpart queries without typos (runs a-d). This finding agrees with previous work~\cite{zhuang2021dealing}, and we stress that it applies also to complex and otherwise highly performing dense retrievers like ANCE and TCT-ColBERTv2. We further note that the use of CharacterBERT alone does not address the problem of robustness to typos: \texttt{CharacterBERT-DR} does now perform significantly better than \texttt{StandardBERT-DR} on MS MARCO dev queries, but the overall effectiveness is poor. 

On the other hand, dense retrievers trained using methods that explicitly address robustness on queries with typos (runs o-r) showcase improved effectiveness. Among those, DRs that use our \texttt{ST} method always perform better than those that use the previous proposed \texttt{Aug} method (except for MRR for \texttt{StandardBERT-DR+ST} on TREC DL 2020 -- difference not significant). Improvements provided on MS MARCO dev queries by our \texttt{ST} compared to \texttt{Aug} for \texttt{StandardBERT-DR} for all metrics and for \texttt{CharacterBERT-DR} for R@1000 are statistically significant; other difference between the two methods across other datasets are at times large, but not significant. We further note that when either of these two forms of training are employed, \texttt{CharacterBERT-DR} always shows higher effectiveness than \texttt{StandardBERT-DR} on queries with typos.
This finding supports our intuition presented in Section~\ref{sec:3}. To further analyse this aspect, in Figure~\ref{fig:droprate-to-cosine} we also plot the MRR drop rate ($\Delta_{MRR}$) and the cosine similarity of \texttt{CharacterBERT-DR+ST} (dashed lines) with respect to tokenization difference; recall that the solid lines refer to \texttt{StandardBERT-DR} (without any specific training tackling queries with typos) and were analysed in Section~\ref{sec:3}. 
Interestingly, our \texttt{CharacterBERT-DR+ST} shows the exact opposite trend of \texttt{StandardBERT-DR}: For queries with typos that resulted in larger tokenization differences (i.e. a larger amount of tokens differ from those obtained by the corresponding query without typos) the $\Delta_{MRR}$ of \texttt{CharacterBERT-DR+ST} decreases, while the encoding similarity between the encoded original queries and the encoded queries with typos increases. This analysis further demonstrates that our method is effective at narrowing the gap (in effectiveness and representation) between the queries with and without typos.

\textit{\textbf{A note on query latency and model size.}} Intuitively, CharacterBERT should result in higher query latency than BERT as it uses Character-CNNs to construct token embeddings, while BERT just uses a lookup table. However, in our experiments, the query latency on GPU of CharacterBERT is just one millisecond slower than that of BERT. This is due to the fact that the number of token embeddings constructed by Character-CNNs is usually smaller than the number of token embeddings constructed by the WordPiece tokenizer. Thus, for the same query, the self-attention computation in the BERT transformer layers is often less time consuming for CharacterBERT than for BERT. Another side effect of Character-CNNs is that CharacterBERT has less model parameters than BERT (105M vs 110M) as it does not need to store the token embeddings in the BERT's WordPiece vocabulary: it only has 262 character embeddings and some extra parameters for the CNNs.

\begin{table}\small
	\resizebox{\columnwidth}{!}{
		\begin{tabular}{|c|l|lll|}
			\hline
			\multirow{2}{*}{Queries}                                                    & \multicolumn{1}{c|}{\multirow{2}{*}{Methods}} & \multicolumn{3}{c|}{DL-typo} \\ \cline{3-5} 
			& \multicolumn{1}{c|}{}                         & nDCG@10   & MRR     & MAP    \\ \hline
			\multirow{9}{*}{\begin{tabular}[c]{@{}c@{}}Without\\ Typos\end{tabular}} 
			& a) \texttt{BM25}                                          & .527      & .633   & .298   \\
			& b) \texttt{ANCE}~\cite{xiong2020approximate}                                          & .606$^a$      & .770$^a$   & .480$^a$   \\
			& c) \texttt{TCT-ColBERTv2}~\cite{lin-etal-2021-batch}                                          & .650$^{ab}$      & \textbf{.890}$^{abi}$   & \textbf{.565}$^{ab}$   \\
			& d) \texttt{StandardBERT-DR}                                      & \textbf{.722}$^{abc}$     & .833$^{a}$   & \textbf{.565}$^{ab}$   \\
			& e) \texttt{CharacterBERT-DR}                                 & .716$^{abc}$      & .855$^{a}$   & .538$^{ab}$   \\ \cline{2-5}
			& f) \texttt{StandardBERT-DR+Aug}~\cite{zhuang2021dealing}                                & \textbf{.737}$^{abc}$      & \textbf{.840}$^{a}$   & \textbf{.594$^{abehi}$}   \\
			& g) \texttt{StandardBERT-DR+ST}                                     & .725$^{abc}$      & .827$^{a}$   & .583$^{ab}$   \\
			& h) \texttt{CharacterBERT-DR+Aug}                       & .713$^{abc}$      & .821$^{a}$   & .539$^{ab}$   \\
			& i) \texttt{CharacterBERT-DR+ST}                            & .706$^{ab}$      & .793$^{a}$   & .539$^{a}$   \\ \hline \bottomrule
			\multirow{9}{*}{\begin{tabular}[c]{@{}c@{}}With\\ Typos\end{tabular}}     
			& j) \texttt{BM25}                                  & .212      & .203   & .104   \\
			& k) \texttt{ANCE}~\cite{xiong2020approximate}                                          & \textbf{.340}$^{j}$      & \textbf{.508}$^{j}$   & \textbf{.245}$^{jm}$   \\
			& l) \texttt{TCT-ColBERTv2}~\cite{lin-etal-2021-batch}                                          & .310$^{j}$      & .421$^{j}$   & .221$^{jm}$   \\
			& m) \texttt{StandardBERT-DR}                                           & .283$^{j}$      & .371$^{j}$   & .167   \\
			& n) \texttt{CharacterBERT-DR}                                 & .297      & .386$^{j}$   & .224$^{j}$   \\ \cline{2-5}
			& o) \texttt{StandardBERT-DR+Aug} ~\cite{zhuang2021dealing}                                & .408$^{jlmn}$      & .502$^{jm}$   & .283$^{jlm}$   \\
			& p) \texttt{StandardBERT-DR+ST}                                      & .433$^{jklmn}$      & .531$^{jlmn}$   & .301$^{jlm}$   \\
			& q) \texttt{CharacterBERT-DR+Aug}                       & .443$^{jklmn}$      & .598$^{jlmn}$   & .326$^{jklmn}$   \\
			& r) \texttt{CharacterBERT-DR+ST}                            & \textbf{.473$^{jklmn}$}      & \textbf{.615$^{jlmn}$}   & \textbf{.348$^{jklmn}$}   \\ \hline
		\end{tabular}
	}
	\caption{DL-typo results. Methods statistically significantly better ($p<0.05$) than others are indicated by superscripts.}
	\label{table:4}
\end{table}

\begin{figure*}[t]
	\begin{minipage}{.4\textwidth}
		\centering \includegraphics[width=0.85\linewidth]{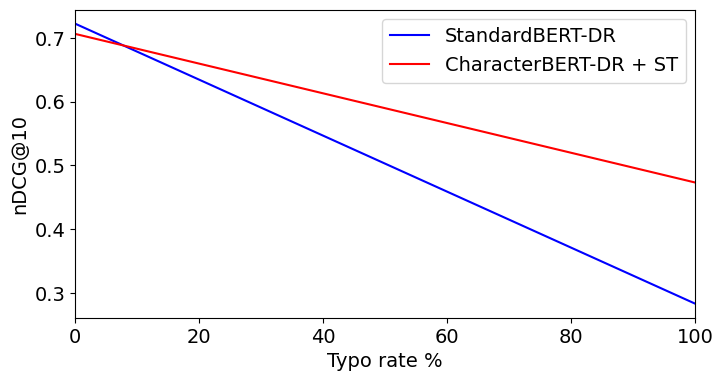}
		\caption{How nDCG@10 changes according to the relative frequency of typos in the query set. \label{fig:effectiveness-to-typorate}}
	\end{minipage} \quad
	\begin{minipage}{.57\textwidth}
		\resizebox{\columnwidth}{!}{
			\begin{tabular}{|c|l|ll|lll|}
				\hline
				\multirow{2}{*}{Queries}                                                    & \multicolumn{1}{c|}{\multirow{2}{*}{Method}}  & \multicolumn{2}{c|}{MS MARCO}                                 & \multicolumn{3}{c|}{DL-typo}                                                    \\ \cline{3-7} 
				& \multicolumn{1}{c|}{}                         & \multicolumn{1}{c}{MRR@10} & \multicolumn{1}{c|}{R@1000} & \multicolumn{1}{c}{nDCG@10} & \multicolumn{1}{c}{MRR} & \multicolumn{1}{c|}{MAP} \\ \hline
				\multirow{5}{*}{\begin{tabular}[c]{@{}c@{}}Without\\ Typos\end{tabular}} 
				& a) \texttt{pyspellchecker} -> \texttt{StandardBERT-DR}             & .276                   & .888                            & .700                         & .811                   & .550                    \\
				& b) \texttt{MSspellchecker} -> \texttt{StandardBERT-DR}             & .324$^{ac}$                        & \textbf{.951}$^{ac}$                               & \textbf{.719}                        & .833                  & \textbf{.563}                     \\
				 & c) \texttt{pyspellchecker} -> \texttt{CharacterBERT-DR}             & .279                       & .887                             & .703                        & .824                   & .527                     \\
				& d) \texttt{MSspellchecker} -> \texttt{CharacterBERT-DR}             & \textbf{.326}$^{ac}$                         & .948$^{ac}$                               & .715                        & \textbf{.855}                   & .538                     \\
				& e) \texttt{CharacterBERT-DR+ST}                         & .325$^{ac}$                      & .950$^{ac}$                               & .706                        & .793                   & .539                    \\ \hline
				\bottomrule
				\multirow{5}{*}{\begin{tabular}[c]{@{}c@{}}With\\ Typos\end{tabular}}     
				& d) \texttt{pyspellchecker} -> \texttt{StandardBERT-DR} & .231                      & .819                             & .475                       & .562                   & .340                     \\
				& f) \texttt{MSspellchecker} -> \texttt{StandardBERT-DR}             & .303$^{dgi}$                   & .920$^{dgi}$                      & \textbf{.716}$^{dgi}$                   & .833$^{dgi}$                 & \textbf{.559}$^{dgi}$                     \\
				& g) \texttt{pyspellchecker} -> \texttt{CharacterBERT-DR}             & .234                       & .821                             & .462                        & .573                   & .339                     \\
				& h) \texttt{MSspellchecker} -> \texttt{CharacterBERT-DR}             & \textbf{.305}$^{dgi}$                         & .930$^{dgi}$                               & .714$^{dgi}$                          & \textbf{.855}$^{dgi}$                     & .539$^{dgi}$                       \\
				& i) \texttt{CharacterBERT-DR+ST}                         & .263$^{dg}$                       & .894$^{dg}$                          & .473                        & .615                   & .348                     \\ \hline
			\end{tabular}
		}
		
		\captionof{table}{Comparison between \texttt{CharacterBERT-DR+ST} and pipelines that involve spell-checkers. Methods statistically significantly better ($p<0.05$) than others are indicated by superscripts.}
		\label{table:5}
	\end{minipage}
	
\end{figure*}

\subsection{Results on Queries with Real Typos}
The results presented so far are obtained with synthetic typo generation. Although, this synthetic evaluation has been used in previous works~\cite{zhuang2021dealing,sun2020adv,penha2021evaluating,wu2021neural}, synthetically generated typos in queries may differ from these encountered in practice.
Our DL-typo dataset allows us to investigate effectiveness in real user queries with typos. Results obtained on DL-typo are reported in Table~\ref{table:4}. Note that runs based on \texttt{StandardBERT-DR} and \texttt{CharacterBERT-DR} (i.e., runs d-i and m-r) were pooled to form our dataset (and thus fully assessed at least up to rank 10), while ANCE and TCT-ColBERTv2 were not pulled: thus their evaluation is affected by a larger number of unjudged passages, and thus direct comparison with the other methods may be unfair\footnote{About 20\% - 30\% passages in the top-10 of ANCE and TCT-ColBERT are unjudged.}.
Overall, results on DL-typo for \texttt{StandardBERT-DR} and \texttt{CharacterBERT-DR} show similar trends to those on the other considered datasets (Table~\ref{table:3}).
In particular, on queries without typos, training with \texttt{Aug} or \texttt{ST} delivers results similar to standard training; however they do improve results on queries with typos -- and our proposed \texttt{CharacterBERT-DR+ST} achieves the best effectiveness across all metrics for queries with typos. While \texttt{CharacterBERT-DR} based models show slightly worse effectiveness than \texttt{standardBERT-DR} for queries without typos, we do not find statistically significant differences, except for MAP for \texttt{standardBERT-DR+Aug}. We further study on this dataset what is the relative percentage of queries with typos that need to be present in a dataset to prefer the use of \texttt{CharacterBERT-DR+ST} over \texttt{StandardBERT-DR}. This analysis is shown in Figure~\ref{fig:effectiveness-to-typorate}: already with only $8\%$ of the overall queries containing typos, \texttt{CharacterBERT-DR+ST} shows higher effectiveness than \texttt{StandardBERT-DR}. This difference becomes statistically significant when about $30\%$ or more of the queries contain typos. We note that several previous studies have observed a high typos rate in search engines query logs~\cite{nordlie1999user,spink2001searching,wang2003mining,wilbur2006spelling,hagen2017large} and fixing those typos could be very expensive\footnote{https://unbxd.com/blog/site-search-and-common-misspellings/}. Our \texttt{CharacterBERT-DR+ST} can be a promising solution for this problem.

\subsection{\texttt{CharacterBERT-DR+ST} vs Spell-checkers}
Next, we compare \texttt{CharacterBERT-DR+ST} with a common pipeline used in search engines for dealing with typos: the pre-processing of queries with spell-checker systems. For this, we employ \texttt{py\-spell\-chec\-ker}\footnote{https://github.com/barrust/pyspellchecker}, which implements a rule-based spell checking algorithm, and the state-of-the-art Microsoft Bing spell checking API (\texttt{MSspellchecker})\footnote{https://docs.microsoft.com/en-us/azure/cognitive-services/bing-spell-check/overview}, which leverages deep learning and statistical machine translation, along with a large amount of crawled data and search engine interactions, to provide accurate and contextual corrections. 
Differently from \texttt{CharacterBERT-DR+ST}, which tackles typos in queries in an end-to-end manner, the use of spell-checkers requires the introduction of an extra step: first it applies the spelling correction algorithm to the query to identify and correct possible typos, and then issues the corrected query to the retrieval model. Empirical comparison is reported with respect to MS MARCO dev queries and DL-typo.


Empirical results are reported in Table~\ref{table:5}. When queries do not contain typos and the MS MARCO dataset is considered, \texttt{Cha\-rac\-terBERT-DR+ST} and the \texttt{MSspellchecker} have similar effectiveness, and both are significantly better than \texttt{pyspellchecker}. In particular, on these queries, \texttt{pyspellchecker} fails because it incorrectly identifies typos in queries where there are not, thus hurting the dense retriever effectiveness. The methods instead do not show statistically significant differences on the queries without typos contained in DL-typo.
When queries with typos are considered, \texttt{MSspellchecker} has a much higher effectiveness than the other methods. However, we note that the \texttt{MSspellchecker} is the most expensive method among those considered in terms of both training and inference, and is likely finetuned across a significantly larger amount of data and labels than the other considered methods. Our \texttt{CharacterBERT-DR+ST} exhibits higher effectiveness than \texttt{pyspellchecker} on both datasets (statistically significant on MS MARCO), with the only exception that nDCG@10 is slightly worse than \texttt{pyspellchecker->StandardBERT-DR} on the DL-typo dataset (no statistical significance), suggesting that our end-to-end method is better than a pipeline with a rule-based spell-checker: it achieves similar, when not higher, effectiveness at a lower computational cost. The use of an end-to-end system like that possible with \texttt{CharacterBERT-DR+ST}  as opposed to a pipeline that includes separate components like a spell-checker and a separate dense retriever, presents a key engineering advantage: a simpler, less complex system, with less components that need to be maintained, monitored and updated.

\section{Conclusions} \label{sec:conclusion}

We extensively analysed the behaviour of BERT-based dense retrievers on queries that contain typos and identified the underlying reasons for which these methods are not robust to typos in queries. Specifically, we unveiled that  BERT's WordPiece tokenizer can dramatically change the input token distributions of the query encoder in presence of typos in the query, thereafter negatively impacting the downstream search task. To overcome this issue, we proposed to use CharacterBERT in combination with a Self-Teaching (ST) adversarial training for learning dense retrievers that are embedding-invariant w.r.t. queries with and without typos.
Our empirical results demonstrated that dense retrievers trained with our proposed CharacterBERT and ST are much more robust to typos in queries, while not deteriorating effectiveness on queries without typos. 
In addition, our methods show the potential for replacing traditional spell-checkers used in retrieval pipelines, thus simplifying the retrieval system and lowering engineering and maintenance operations. However, we also observe there are still gaps compared to more sophisticated and highly trained deep learning based spell-checkers, highlighting that further improvements to the proposed methods are required to obtain an end-to-end dense retriever that can replace search engine pipelines that use these more complex spell-checkers. Furthermore, we also observe that the application of the proposed CharacterBERT and ST to more complex training approaches such as TCT-ColBERTv2 (not investigated here) may well lead to even stronger, typo-robust dense retrievers. 
Code, experimental results and DL-typo dataset are publicly available at \url{https://github.com/ielab/CharacterBERT-DR}.

\begin{acks}
This research is partially funded by the Grain Research and Development Corporation project AgAsk (UOQ2003- 009RTX). We thank Ahmed Mourad, Harry Scells, Shuai Wang from UQ ielab and Yun Wen for helping assembling the DL-typo dataset. 
\end{acks}

\bibliographystyle{ACM-Reference-Format}
\bibliography{sigir2022-typo-distillation}

\appendix

\end{document}